# Superconductivity in BiS$_2$-Based Layered Compounds

Yoshikazu Mizuguchi[a],*

[a]*Tokyo Metropolitan University, 1-1, Minami-osawa, Hachioji, 192-0397, Japan*

**Abstract**

Crystal structure and physical properties of the novel BiS$_2$-based layered superconductors are briefly reviewed. Superconductivity in the BiS$_2$-based layered compounds is induced by electron doping into the BiS$_2$ conduction layers. The superconducting properties seem to correlate with the crystal structure. Possible strategies for increasing transition temperature in this family are discussed.

**Introduction**

Since the discovery of the cuprate layered superconductors, materials possessing a layered crystal structure have been one of the mostly-studied systems on exploration of new superconductors [1]. The discovery of the Fe-based superconductors in 2008 has also accelerated studies on new layered materials [2]. One of the reasons for the tremendous amount of attentions in the layered superconductors is the observation of unconventional paring mechanisms due to low-dimensional characteristics of conductive electrons. Another merit of layered materials in exploring for new superconductors is the variation of crystal structure. In general, the layered superconductors possess a crystal structure composed of an alternate stacking of a common superconducting layer and a blocking (spacer) layer. In the Fe-based family, the Fe$_2$An$_2$ layers (An = P, As, S, Se or Te) could act as a common superconducting layer, and a lot of Fe-based superconductors have been discovered by changing the blocking layer structure [2-4]. Namely, if we discover a new type of superconducting layers, we can design a lot of layered superconductors. In 2012, the novel layered superconductors with a BiS$_2$-based superconducting layer have been discovered [5,6]. So far, 11 superconductors have been discovered, and some notable characteristics have been observed in this family [7-11]. Here, the crystal structure and physical properties of the BiS$_2$-based superconductors are briefly reviewed.

**Crystal structure and superconducting properties**

Figure 1(a) and (b) show the schematic images of the crystal structure of typical BiS$_2$-based superconductor LaOBiS$_2$ and Bi$_4$O$_4$S$_3$, respectively. Both materials are composed of an alternate stacking of the BiS$_2$ double layers and the blocking layer. Electron carriers, which are essential for the appearance of superconductivity in the BiS$_2$-based family, can be generated (controlled) by modifying the structure and the composition at the blocking layers. In the REOBiS$_2$ system (RE = Rare earth), electron carriers can be generated upon a partial substitution of O$^{2-}$ by F$^-$ [6], which is an electron-doping strategy used in the FeAs-based superconductors. In the Bi$_4$O$_4$S$_3$ superconductor, partial defects at the (SO$_4$)$^{2-}$ site can provide electron carriers into the BiS$_2$ layers [5].

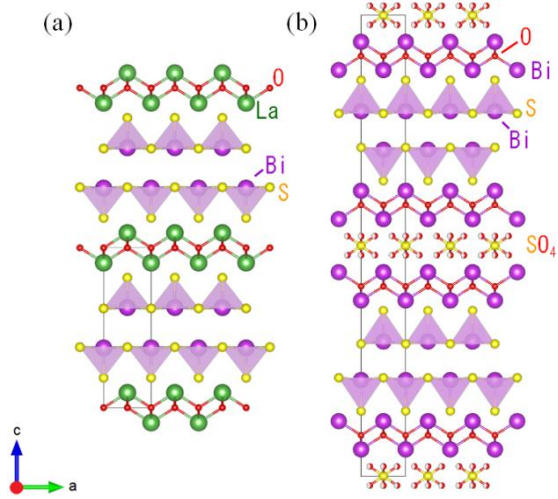

Fig. 1. (a) Crystal structure of LaOBiS$_2$. (b) Crystal structure of Bi$_4$O$_4$S$_3$. In this figure, the structure is depicted with full site occupancy: Namely, the elemental composition of the shown structure is Bi$_6$O$_8$S$_5$. In real, there would be ~50 % defects at the (SO$_4$)$^{2-}$ site, which corresponds to the composition of Bi$_4$O$_4$S$_3$.

The physical properties of BiS$_2$-based compounds are introduced with the data of the LaO$_{1-x}$F$_x$BiS$_2$ system. Figure 2(a) displays the temperature dependence of resistivity for LaOBiS$_2$ (parent phase) and LaO$_{0.5}$F$_{0.5}$BiS$_2$. The parent phase exhibits semiconducting transport behavior. With the partial substitution of O by F, the resistivity decreases and superconducting transition is observed at ~3 K for LaO$_{0.5}$F$_{0.5}$BiS$_2$. The superconducting properties can be enhanced by annealing the LaO$_{0.5}$F$_{0.5}$BiS$_2$ superconducting sample under high pressure (HP). Figure 2(b) shows the temperature dependence of resistivity for as-grown and HP-annealed LaO$_{0.5}$F$_{0.5}$BiS$_2$. The onset of the transition temperature ($T_c$) is clearly enhanced from 3 to 10.6 K by the HP annealing [6,10]. It was found that the uniaxial strain (contraction) along the $c$ axis is generated by the HP annealing [11]. These facts indicate that the optimization local structure is essential for the appearance of bulk superconductivity at a higher temperature in the BiS$_2$-based superconductors. It was also reported that the application of external pressure could enhance the $T_c$ of as-grown LaO$_{0.5}$F$_{0.5}$BiS$_2$. Furthermore, the crystal structure analysis under HP showed that the structural transition from tetragonal to monoclinic was positively linked with the enhancement of $T_c$ under high pressure [12]. Theoretical investigations suggested that the slight changes in the local crystal structure such as the $z$ coordinate of S at the BiS$_2$ layer could largely affect the band structure [13,14]. In these regards, the correlation between superconductivity and local crystal structure should be important to understand the mechanisms and to enhance the $T_c$ in the BiS$_2$-based superconductors.

Figure 3(a) shows the temperature dependence of normalized resistivity around superconducting transition for the typical BiS$_2$-based superconductors, LaO$_{0.5}$F$_{0.5}$BiS$_2$ (HP-annealed), CeO$_{0.3}$F$_{0.7}$BiS$_2$ (HP-annealed), NdO$_{0.7}$F$_{0.3}$BiS$_2$ (as-grown) and Bi$_4$O$_4$S$_3$ (as-grown). Here the transport data for the samples with the highest $T_c$ in the respective systems are summarized. The midpoints of $T_c$ are summarized in Fig. 3(b). To discuss the tendency of the $T_c$, the area where the superconducting properties ($T_c$) are optimized is highlighted with a green square. The optimized $T_c$ in the LaOBiS$_2$ system is the highest. Then, the optimized $T_c$ tends to decrease by changing the blocking layer structure in an order of La$_2$O$_2$, Ce$_2$O$_2$, Nd$_2$O$_2$ and Bi$_4$O$_4$(SO$_4$). The tendency can be understand with the change in the ionic radius of the anion (La$^{3+}$, Ce$^{3+}$, Nd$^{3+}$ and Bi$^{3+}$) in the blocking layer. The ionic radius of the anion directly tune the $a$ axis length in this family. In fact, the optimized $T_c$ in the BiS$_2$-based superconductors well correlates with the length of the $a$ axis.

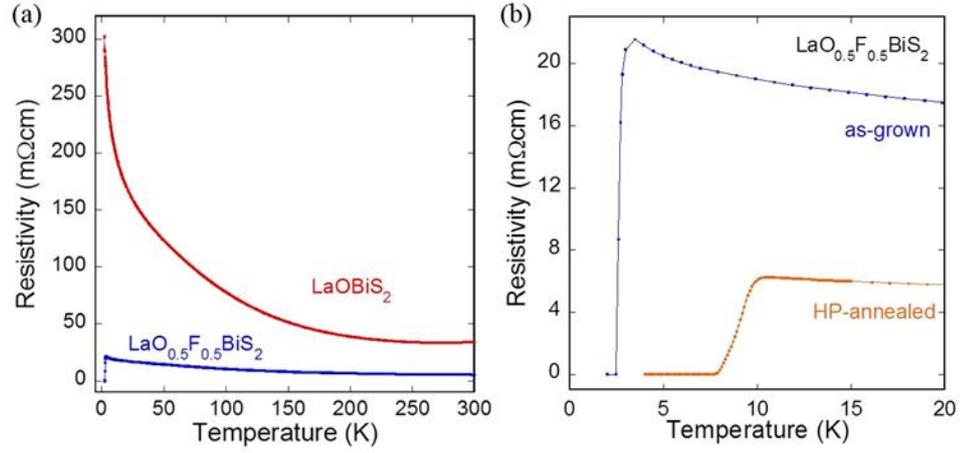

Fig. 2. (a) Temperature dependence of resistivity for $LaOBiS_2$ and $LaO_{0.5}F_{0.5}BiS_2$. (b) Temperature dependence of resistivity for as-grown and HP-annealed $LaO_{0.5}F_{0.5}BiS_2$.

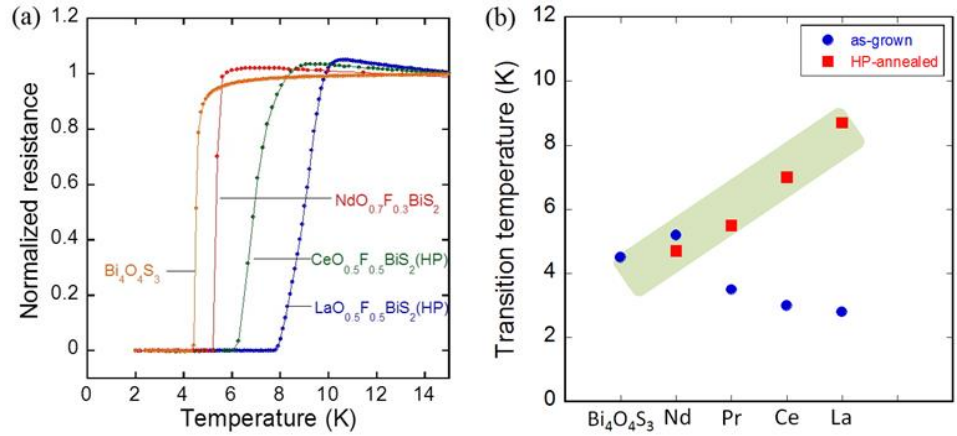

Fig. 3. (a) Temperature dependence of normalized resistivity around superconducting transition for the typical $BiS_2$-based superconductors, $LaO_{0.5}F_{0.5}BiS_2$ (HP-annealed), $CeO_{0.3}F_{0.7}BiS_2$ (HP-annealed), $NdO_{0.7}F_{0.3}BiS_2$ (as-grown) and $Bi_4O_4S_3$ (as-grown). (b) Transition temperatures (midpoint) are summarized.

**Strategies to enhance superconducting properties**

In the above, possible two structural parameters are suggested to be essential for superconductivity in the $BiS_2$ family. The first one is the local structure correlated with the uniaxial compression along the *a* axis, which can be optimized by applying high pressure. This should be essential for the appearance of bulk superconductivity in the respective systems. The other parameter is the length of the *a* axis, which can be tuned by changing the blocking layer. Unfortunately, the largest rare earth is La; hence, we cannot synthesize new $REOBiS_2$ materials with larger *a* axis. Therefore, we should discover new blocking layers. For example, the structure similar to $Bi_6O_8S_5$, Fig. 1(b), is a candidate. If the Bi could be substituted with rare earth, larger *a* axis could be obtained. Another strategy is to use a perovskite-oxide blocking layer, which is formed in the Fe-based superconductor $Sr_4Sc_2O_6Fe_2P_2$ [17].

Recently, superconductors with Bi-Se and Bi-Te conduction layers were discovered. Maziopa et al. reported that $LaO_{0.5}F_{0.5}BiSe_2$ with a structure similar to $LaOBiS_2$ showed superconductivity below 2.6 K [18,19]. Malliakas et al. reported that $CsBi_4Te_6$ exhibited superconductivity below 4.4 K by electron carrier doping. The Bi-Te blocks of $CsBi_4Te_6$ are similar to the NaCl structure as well as the $BiS_2$ layers. Therefore, superconductivity observed in doped $CsBi_4Te_6$ could be driven with the mechanisms same as the $BiS_2$-based superconductors. Hence, replacement of $BiS_2$ layers by Bi-Se or Bi-Te layers could be a promising strategy for enhancing $T_c$ of Bi-based superconductors.

**Acknowledgement**

This work was partly supported by a Grant-in-Aid for Scientific Research for young scientists (A).

**References**

[1] Bednorz JB, Muller K. Possible high $T_c$ superconductivity in the Ba− La− Cu− O system. Z. Phys. B 1986;64:189-193.

[2] Kamihara Y, Watanabe T, Hirano M, Hosono H. Iron-Based Layered Superconductor $LaO_{1-x}F_xFeAs$ ($x$ = 0.05 - 0.12) with $T_c$ = 26 K. J. Am. Chem. Soc. 2008;130:3296-3297.

[3] Mizuguchi Y, Fujihisa H, Gotoh Y, Suzuki K, Usui H, Kuroki K, Demura S, Takano Y, Izawa H, Miura O. $BiS_2$-based layered superconductor $Bi_4O_4S_3$. Phys. Rev. B 2012;86:220510(1-5).

[4] Mizuguchi Y, Demura S, Deguchi K, Takano Y, Fujihisa H, Gotoh Y, Izawa H, Miura O. Superconductivity in novel $BiS_2$-based layered superconductor $LaO_{1-x}F_xBiS_2$. J. Phys. Soc. Jpn. 2012;81:114725(1-5).

[5] Demura S, Mizuguchi Y, Deguchi K, Okazaki H, Hara H, Watanabe T, Denholme SJ, Fujioka M, Ozaki T, Fujihisa H, Gotoh Y, Miura O, Yamaguchi T, Takeya H, Takano Y. New Member of $BiS_2$-Based Superconductor $NdO_{1-x}F_xBiS_2$. J. Phys. Soc. Jpn. 2013;82:033708(1-3).

[6] Xing J, Li S, Ding X, Yang H, Wen HH. Superconductivity appears in the vicinity of semiconducting-like behavior in $CeO_{1-x}F_xBiS_2$. Phys. Rev. B 2012;86:214518(1-5).

[7] Jha R, Kumar A, Singh SK, Awana VPS. Synthesis and Superconductivity of New $BiS_2$ Based Superconductor $PrO_{0.5}F_{0.5}BiS_2$. J. Sup. Novel Mag. 2013;26:499-502.

[8] Yazici D, Huang K, White BD, Chang AH, Friedman AJ, Maple MB. Superconductivity of F-substituted $LnOBiS_2$ (Ln = La, Ce, Pr, Nd, Yb) compounds. Philosophical Magazine 2012;93:673-680.

[9] Yazici D, Huang K, White BD, Jeon I, Burnett VW, Friedman AJ, Lum IK, Nallaiyan M, Spagna S, Maple MB. Superconductivity induced by electron doping in $La_{1-x}M_xOBiS_2$ (M = Ti, Zr, Hf, Th). Phys. Rev. B 2013;87:174512(1-8).

[10] Deguchi K, Mizuguchi Y, Demura S, Hara H, Watanabe T, Denholme SJ, Fujioka M, Okazaki H, Ozaki T, Takeya H, Yamaguchi T, Miura O, Takano Y. Evolution of superconductivity in $LaO_{1-x}F_xBiS_2$ prepared by high-pressure technique. EPL 2013;101:17004(p1-p5).

[11] Kajitani J, Deguchi K, Omachi A, Hiroi T, Takano Y, Takatsu H, Kadowaki H, Miura O, Mizuguchi Y. Correlation between crystal structure and superconductivity in $LaO_{0.5}F_{0.5}BiS_2$. arXiv:1306.3346.

[12] Tomita T, Ebata M, Soeda H, Takahashi H, Fujihisa H, Gotoh Y, Mizuguchi Y, Izawa H, Miura O, Demura S, Deguchi K, Takano Y. Pressure-induced Enhancement of Superconductivity in $BiS_2$-layered $LaO_{1-x}F_xBiS_2$. arXiv:1309.4250.

[13] Usui H, Suzuki K, Kuroki K. Minimal electronic models for superconducting $BiS_2$ layers. Phys. Rev. B 2012;86:220501(1-5).

[14] Suzuki K, Usui H, Kuroki K. Minimum model and its theoretical analysis for superconducting materials with $BiS_2$ layers. Phys. Procedia 2013;45:21-24.

[15] Demura S, Deguchi K, Sato K, Honjyo R, Mizuguchi Y, Okazaki H, Hara H, Watanabe T, Denholme SJ, Fujioka M, Ozaki T, Miura O, Yamaguchi T, Takeya H, TakanoY. Coexistence of bulk superconductivity and ferromagnetism in $CeO_{1-x}F_xBiS_2$. to appear.

[16] Kajitani J, Deguchi K, Mizuguchi Y, Takano Y et al. Superconducting properties of HP-annealed $PrO_{0.5}F_{0.5}BiS_2$. to appear.

[17] Ogino H, Matsumura Y, Katsura Y, Ushiyama K, Horii S, Kishio K, Shimoyama J. Superconductivity at 17 K in $(Fe_2P_2)(Sr_4Sc_2O_6)$: a new superconducting layered pnictide oxide with a thick perovskite oxide layer. Supercond. Sci. Technol. 2009;22:075008.

[18] Maziopa AK, Guguchia Z, Pomjakushina E, Pomjakushin V, Khasanov R, Luetkens H, Biswas PK, Amato A, Keller H, Conder K. Superconductivity in a new layered bismuth oxyselenide: $LaO_{0.5}F_{0.5}BiSe_2$. arXiv:1310.8131.

[19] Malliakas CD, Chung DY, Claus H, Kanatzidis MG. Superconductivity in the Narrow-Gap Semiconductor $CsBi_4Te_6$. J. Am. Chem. Soc. 2013;130:14540-14543.